\documentclass[doublecol]{epl2} 

\usepackage{graphicx}
\usepackage{dcolumn}
\usepackage{bm}
\usepackage{amssymb}

\title{Deformation of Equilibrium Shape of a Vesicle Induced by Injected Flexible Polymers}
\shorttitle{Title} 

\author{Yutaka Oya, Katsuhiko Sato \and Toshihiro Kawakatsu}
\shortauthor{Y. Oya \etal}

\institute{Department of Physics, Tohoku University, Sendai, 980-8578, Japan}

\abstract{
Using field theoretic approach, we study equilibrium shape deformation of a vesicle induced by the presence of enclosed flexible polymers, which is a simple model of drug delivery system or endocytosis.  
To evaluate the total free energy of this system, it is necessary to calculate the bending elastic energy of the membrane, the conformation entropy of the polymers and their interactions.  For this purpose, we combine phase field theory for the membrane and self-consistent field theory for the polymers.  
Simulations on this coupled model system for axiosymmetric shapes show a shape deformation of the vesicle induced by introducing polymers into it.  We examined the dependence of the stability of the vesicle shape on the chain length of the polymers and the packing ratio of the vesicle.  We present a simple model calculation that shows the relative stability of the prolate shape compared to the oblate shape.
}

\begin{document}

\maketitle

Micelles and vesicles are closed forms of membranes that are composed of amphiphilic molecules, such as surfactants or lipid molecules.  These molecules are not only elementary components of biological cells\cite{Cell} but also important materials in industrial sciences, for example, surface coating, oil recovery, cosmetics and so on\cite{Banat}.
Among various functions and applications of vesicles, a vesicle that encloses polymers can be used as a simple model of drug delivery system (DDS) and the endocytosis in biological cells\cite{Cell}.  In these phenomena, shape deformation, fusion, and fission of vesicles induced by the enclosed polymers are essential to our understanding the total process.

Nakaya {\it et al.} performed a small angle neutron scattering experiment on an inverted micellar phase of a surfactant solution where hydrophilic polymers are enclosed in these micelles\cite{Nakaya}.  They reported that micelles show anisotropic deformation upon the inclusion of the polymers inside them.  In this process, there are several candidates for the reason of the transition, i.e. the configuration entropy of the centers of mass of the enclosed polymers and the solvent and the conformation entropy of the polymers.

In the present paper, we introduce a new theoretical model of a closed-form membrane (a vesicle) that contains polymers, and clarify the mechanism of its shape deformation.  A standard technique to study such phenomena is a molecular simulation where the polymers are modeled by bead-spring chains and the membrane is modeled either by a set of short bead-spring chains composed of hydrophilic and hydrophobic beads\cite{MolecularModelMembrane,Urakami} or by a set of vertex points of a triangular mesh on the membrane surface (so-called surface element method)\cite{MeshModelMembrane}.  One of the difficulties of these techniques is that an evaluation of the free energy of the system is not easy for such molecular simulations\cite{AllenTildesley}.
To overcome this difficulty, we combine the self-consistent field (SCF) theory for polymers\cite{SCF, MatsenSchick} and the phase field (PF) theory for membranes\cite{Du, Campelo}.
In the SCF theory, the probability distribution of the conformation of polymer chains is evaluated in terms of path integral $Q(0, {\bf r}_0; N, {\bf r}_N)$, which corresponds to the statistical weight of a polymer chain composed of $N+1$ segments whose two end segments (denoted by indices 0 and $N$) are at ${\bf r}_0$ and ${\bf r}_N$, respectively\cite{SCF}.  On the other hand, in the PF theory, the membrane is described with a scalar field $\psi({\bf{r}})$, where $\psi({\bf r}) = 0$, $\psi({\bf r}) > 0$ and $\psi({\bf r}) < 0$ correspond to the membrane surface, inside and outside regions of the membrane, respectively.  It should be noted that in the PF theory the membrane is treated as a curved surface with a finite thickness.  This treatment is different from that of the usual Helfrich's bending elastic model where the thickness of the membrane is assumed to be negligible\cite{Helfrich,Seifert}.

Let us describe the detail of our model vesicle that contains polymers.  The target system is a three component mixture composed of amphiphilic molecules (membrane of the vesicle), polymers, and the solvent.  For simplicity, we assume that an amphiphilic molecule, a solvent molecule and a polymer segment have the same volume.  In our model, all of these three components are described in terms of their local density distributions denoted as $\varphi_{\rm M}({\bf r})$, $\varphi_{\rm P}({\bf r})$ and $\varphi_{\rm S}({\bf r})$, respectively.  We require an incompressibility condition on these three components;
\begin{equation}
  \varphi_{\rm M}({\bf r}) + \varphi_{\rm P}({\bf r}) 
  + \varphi_{\rm S}({\bf r}) = 1,
\label{Incompressibility}
\end{equation}
where $\varphi_{\rm M}({\bf r})$ is related to the phase field $\psi({\bf r})$.  This incompressibility condition produces the coupling between the PF and the SCF.

Using these field variables, the total free energy $F_{\rm total}$ of the combined system of the membrane, polymers and the solvent multiplied by $\beta = 1/k_{\rm B}T$ is given in the following form:
\begin{eqnarray}
    \beta F_{\rm total} 
        &=& \beta F_{\rm PF}[\psi] 
         + \sigma \left( {\cal A}_{\rm total}-{\cal A}_{\rm total}^{(0)} \right)
         + \mu \left( {\cal V}_{\rm in} - {\cal V}_{\rm in}^{(0)} \right)
                                                                          \nonumber \\
        &+& \beta F_{\rm SCF}[\psi,\{ \varphi_{K} \}].
\label{TotalFreeEnergy}
\end{eqnarray}

The first term on the right-hand side of eq.~(\ref{TotalFreeEnergy}) is the free energy of the membrane obtained with the PF theory.  
In a non-dimensional form, it is given as\cite{Du_PrivateCommunication}
\begin{equation}
    \beta F_{\rm PF} 
                  = \frac{3 \kappa}{4 \sqrt{2}} \int d{\bf{r}}
                    \{ -\psi({\bf{r}}) + \psi({\bf{r}})^{3} 
                       - \nabla^{2}\psi({\bf{r}}) \}^{2},
\label{PF_Free_Energy} 
\end{equation}
where $\kappa$ is the bending elastic modulus and the unit of length is chosen as the membrane thickness.  In eq.~(\ref{PF_Free_Energy}), we neglected the effects of the spontaneous curvature and the Gaussian curvature for simplicity.

The surface area element of the membrane is given by\cite{Du,Du_PrivateCommunication}
\begin{equation}
    {\cal A}[\psi({\bf{r}})] = \frac{3 \sqrt{\mathstrut 2}}{4}
                                 \left[ \frac{1}{2}\mid \nabla
                                 \psi({\bf{r}}) \mid^{2} 
                               + \frac{1}{4} 
                               ( \psi ({\bf{r}})^{2} - 1 )^{2} \right].
\label{AreaElement}
\end{equation}
Since each amphiphilic molecule occupies a certain constant area on the membrane surface, the local surface area of the membrane is proportional to the local number density of the amphiphilic molecules, $\varphi_{\rm M}({\bf{r}})$.  Therefore, the following relation holds;
\begin{equation}
  \varphi_{\rm M}({\bf{r}}) = C {\cal A} [ \psi({\bf{r}}) ],
\end{equation}
where $C$ is a normalization constant that should be determined so that $\varphi_{\rm M}({\bf{r}})$ takes its maximum value 1 on the central surface of the membrane.

The total area of the membrane surface ${\cal A}_{\rm total}$ and the total enclosed volume by the membrane ${\cal V}_{\rm in}$ are given by
\begin{eqnarray}
  {\cal A}_{\rm total}[\psi({\bf r})] 
     & = & \int {\cal A}[\psi({\bf r})] d{\bf r},
                                                   \nonumber \\
                                                   \nonumber \\
  {\cal V}_{\rm in}[\psi({\bf r})] 
                     & = & \int_{\psi({\bf r}) > 0}
                      \left( 1 - \varphi_{\rm M}({\bf r}) \right) d{\bf r}, 
\end{eqnarray}
respectively.  
In the simulations, we fix these quantities ${\cal A}_{\rm total}$ and ${\cal V}_{\rm in}$ to given values ${\cal A}_{\rm total}^{(0)}$ and ${\cal V}_{\rm in}^{(0)}$ by using Lagrange multipliers $\sigma$ and $\mu$ as are described in the second and third terms on the right-hand side of eq.~(\ref{TotalFreeEnergy})\cite{VolumeConstraint}.  These constraints mean that both of the exchange of amphiphilic molecules between the membrane and the environment and the permeation of solvent across the membrane are very slow compared to the equilibration time scales of the membrane shape and the polymer conformations.

The last term on the right-hand side of eq.~(\ref{TotalFreeEnergy}) is the free energy of the polymers and the solvent calculated with the SCF theory.
In the SCF calculation, we describe the polymer chains using the path integral $Q(0, {\bf r}_0; N, {\bf r}_N)$ while the solvent molecules are assumed to be point particles which posses only translational degrees of freedom. 

The mean field potential used in the SCF calculation is assumed to have the form
\begin{equation}
    V_{K} = \sum_{K^{\prime }} \chi_{KK^{\prime }}
             \varphi_{K^{\prime}}({\bf{r}}) + \gamma({\bf{r}}),
\end{equation}
where indices $K$ and $K^{\prime}$ represent either the polymer (P) or the solvent (S), and $\chi_{KK^{\prime }}$ is Flory interaction parameter between the segments of $K$ and $K^{\prime}$ types, and $\gamma({\bf r})$ is the Lagrange multiplier for the incompressible condition eq.~(\ref{Incompressibility}), which produces a coupling between the polymer chains and the membrane.

For simplicity, we assume that all the interaction parameters $\chi_{KK^{\prime}}$ vanish except for $\chi_{PS} \equiv \chi$. 
We also assume that all polymer chains are confined in the vesicle while the solvent fills both inside and outside regions of the vesicle.  (Due to the constraint of the fixed enclosed volume ${\cal V}_{\rm in}$, there is essentially no exchange of solvent across the membrane.)
To realize this condition, we set the path integral for polymers to be zero outside the vesicle ($Q(0, {\bf r}_0; i, {\bf r}) = 0$ for $\psi({\bf r}) < 0$), while we impose no restrictions on the region of the distribution of the solvent.  
As a result of this SCF calculation, we obtain the following contributions from the SCF part to the total free energy\cite{SCF}
\begin{equation}
  \beta F_{\rm SCF}[\psi, \{ \varphi_K \}]]
    \equiv \beta \left( 
                      F_{\rm P} + F_{\rm S} + F_{\rm int} + F_{\rm incomp}
                  \right),
\end{equation}
where $F_{\rm P}$, $F_{\rm S}$, $F_{\rm int}$ and $F_{\rm incomp}$ are contributions from polymers, solvent, segment interactions, and incompressibility, respectively.  These components are defined as follows;
\begin{eqnarray} 
  && \beta F_{\rm P}
          = - M_{\rm P} \ln \int d{\bf r}_0 \int d{\bf r}_N
                                Q_{\rm  P}(0, {\bf r}_0; N, {\bf r}_N)
                                                       \nonumber \\
  && ~~~~~~~   - \int d{\bf r} V_{\rm P}({\bf r}) \varphi_{\rm P}({\bf r})
               + M_{\rm P} \ln M_{\rm P} - M_{\rm P},   
                                                       \nonumber \\
  && \beta F_{\rm S} 
          = \int d{\bf r} 
             \left[ 
                \varphi_{\rm S}({\bf r}) \ln \varphi_{\rm S}({\bf r})
                - \varphi_{\rm S}({\bf r})
             \right],
                                                       \nonumber \\
  && \beta F_{\rm int} 
       = \frac{1}{2} \sum_{KK^{\prime}} \chi_{KK^{\prime}}
               \int d{\bf r} \varphi_K({\bf r}) \varphi_{K^{\prime}}({\bf r}),
                                                       \nonumber \\
  && \beta F_{\rm incomp}
       = \int \gamma({\bf r}) 
               \left\{ \varphi_{\rm P}({\bf r}) + \varphi_{\rm S}({\bf r})
                    + \varphi_{\rm M}({\bf r}) - 1
               \right\} d {\bf r},
                                                        \nonumber \\
\end{eqnarray}
where $M_{\rm P}$ and $M_{\rm S}$ are total numbers of polymers and solvent molecules in the system.

There are three important non-dimensional parameters that specify the state of the system.  These are the ``packing ratio" inside the vesicle $v$, the volume fraction of polymers inside the vesicle $\phi$, and the chain length of the polymer $N$.  
The packing ratio $v$ is defined as a ratio between the actual enclosed volume ${\cal V}_{\rm in}^{(0)}$ and that of the spherical vesicle with the same surface area ${\cal A}_{\rm total}^{(0)}$\cite{Seifert,Du}:
\begin{equation}
    v = {\cal V}_{\rm in}^{(0)} 
         \left/
         \left[
            \frac{\displaystyle 4 \pi}{\displaystyle 3} 
                 \left( 
                    \frac{\displaystyle {\cal A}_{\rm total}^{(0)}}
                          {\displaystyle 4 \pi}
                 \right)^{3/2}
        \right]
        \right. .
\label{PackingFraction}
\end{equation}
Due to the existence of the solvent in our system, the stable regions of $v$ for oblate and prolate shapes in the absence of polymers are somewhat different from those reported in Ref.~\cite{Du}.  In our simulations, we found that the prolate shape is always stable than the oblate shape for $v < 0.7$ when the membrane does not contain polymers. 
By changing the parameters $v$, $\phi$ and $N$, we examined the stable shape of the vesicle.

\begin{figure}[t]
   \begin{center}
   \includegraphics[width=40mm]{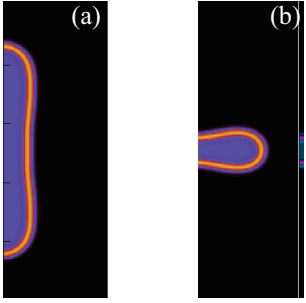}
\caption{\label{VesicleShapeChi0} Typical vesicle shapes: (a) the prolate and (b) the oblate shapes are shown for the case with $v=0.5$, $\phi=0.1$, $\chi = 0.0$, and $N=100$, respectively.  The distributions of the polymers inside the vesicle are also shown by shading.}
  \end{center}
\end{figure}

Assuming axiosymmetric shapes, we minimized the free energy eq.(\ref{TotalFreeEnergy}) with respect to $\{ \varphi_K({\bf r}) \}$ and $\psi({\bf r})$ by an iteration method.  We used a cylindrical coordinate system with $256 \times 80$ mesh points in axial and radial directions, respectively, with mesh width $\Delta x = 0.5$.
Figures~\ref{VesicleShapeChi0}(a) and (b) show two typical shapes of the membrane, {\it i.e.,} the prolate and the oblate shapes, respectively, obtained for the case with $v = 0.5$, $\phi = 0.1$, $\chi = 0.0$, and $N=100$.  The left edge of the figure is the axis of revolution.

\begin{figure}[htbp]
   \begin{center}
   \includegraphics[width=60mm]{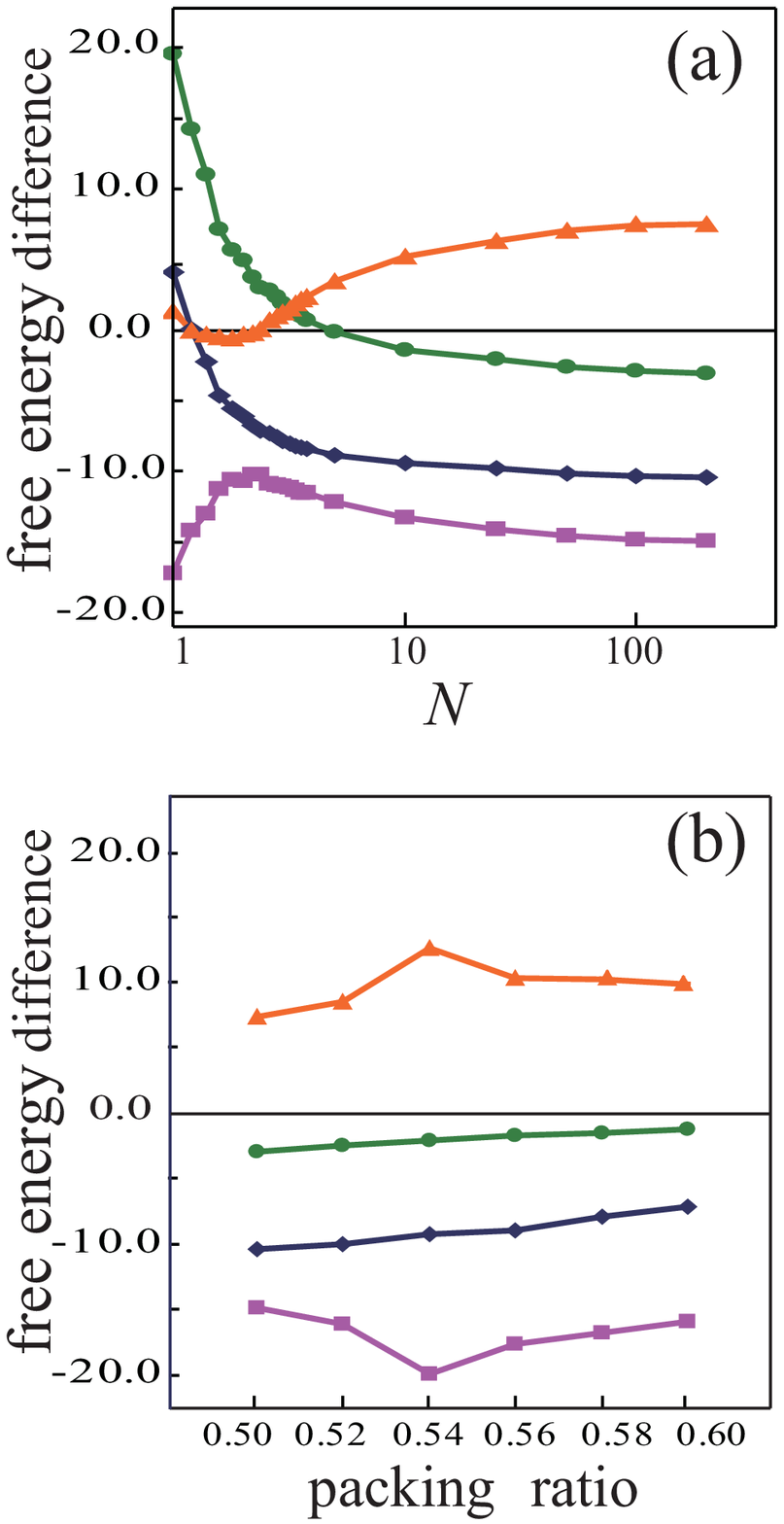}
   \caption{\label{FreeEnergyDifference} Each components of the total free energy, $\beta F_{\rm PF}$ ($\blacksquare$), $\beta \left( F_{\rm P} + F_{\rm incomp} \right)$ ($\bullet$), $\beta F_{\rm S}$ ($\blacktriangle$), and $\beta F_{\rm total}$ ($\blacklozenge$) are shown.  (a) Dependence on the chain length $N$ for the case with $v=0.5$, $\chi=0.0$ and $\phi = 0.1$, and (b) the dependence on the packing ratio $v$ defined in eq.~(\ref{PackingFraction}) for the case with $\chi=0.0$, $\phi = 0.3$, and $N=100$, respectively.  
  Note that $\beta F_{\rm int} = 0$ because $\chi = 0$ for both figures.  For each component, $\beta F({\rm prolate}) - \beta F({\rm oblate})$ is shown.}
   \end{center}
\end{figure}

In Fig.~\ref{FreeEnergyDifference}, we show the dependences of the components of the free energy on the chain length $N$ for the athermal case with $v = 0.5$, $\phi = 0.1$ and $\chi = 0.0$ ({\it i.e.} $F_{\rm int} = 0$).  Shown are $\beta F_{\rm PF}$, $\beta (F_{\rm P} + F_{\rm incomp})$, $\beta F_{\rm S}$, and the total free energy $\beta F_{\rm total}$ defined by eq.(\ref{TotalFreeEnergy}), respectively.  These quantities are the difference between the values for the prolate vesicle and that for the oblate vesicle.  Thus, a negative value means that the prolate shape has lower free energy than the oblate case.
In this figure, the constraint terms that include $\sigma$ and $\mu$ are included in $F_{\rm PF}$ although they give only negligible contributions. 

Figure~\ref{FreeEnergyDifference}(a) indicates that both the conformation entropy of polymers and the bending elastic energy of the membrane tend to prefer prolate shape when the chain length is increased.  On the other hand, the contribution from the translational entropy of the solvents shows a more complex behavior.  In the short chain length region, this contribution once decreases and then it turns to increase when the chain length becomes longer.  As a sum of these components, the total free energy difference decreases monotonically as the polymer chain length is increased, leading to the equilibrium prolate shape in the long chain region.

The complex behavior of the translational entropy of the solvent is understood considering the effect of the depletion layer of the polymers near the membrane.  When the polymer chain length becomes comparable to the membrane thickness (around $N \sim 3$), the width of the depletion layer is negligibly thin, and the solvent distributes almost uniformly inside the vesicle, which maximize the translational entropy of the solvent.  As the chain length is increased, a clear depletion layer is formed, and the solvent distribution inside the vesicle becomes inhomogeneous, which cause a decrease of the translational entropy of the solvent molecules ({\it i.e.} an increase in the free energy).

The opposite behavior of this translational entropy of the solvent in the very short chain length region ($N < 3$) is an artifact of the present phase field modeling of membrane which has a finite thickness that is the same order as the gyration radius of the polymer with $N \sim 5$.  Due to the smooth density profile and the finite thickness of the membrane distribution $\varphi_M({\bf r})$, either the polymer segments or the solvent molecules must come into the membrane region to fill the vacancy.  Such invading molecules are strongly repelled by the membrane and cause an increase in the free energy.  As the rate of this increase is different for the prolate and the oblate shapes, it leads to the steep increase of $\beta F_{\rm S}$ for $N < 3$.

To understand why the conformation entropy of polymers $\beta F_{\rm P}$ prefers prolate shape, we give a simple interpretation. 
Let us approximate an oblate or a prolate shape with a cylinder with diameter $x$ and height $y$.  These two values are determined when the total surface area and the enclosed volume are given.  These conditions lead to $x^2 y = C_1$ and $x^2 + 2 x y = C_2$, where $C_1$ and $C_2$ are constants that correspond to the total enclosed volume multiplied by $4/\pi$ and the total surface area multiplied by $2/\pi$, respectively.  
Solving these set of equations for given $C_1$ and $C_2$ gives 3 solutions $(x_i, y_i)$ ($i = 1, 2, 3$) where $x_3 < 0 < x_2 < x_1$.  Obviously, the solution $x_3 <0$ is unphysical.  The other two solutions correspond to the oblate $(x_1, y_1)$ and prolate $(x_2, y_2)$, respectively.  If $x_2$ is small, we obtain up to the first order in $x_2$ that $x_2 = 2 C_1/C_2 \equiv L_{\rm pr}$ (prolate) and $y_1 = C_1/(2 C_2) \equiv L_{\rm ob}$ (oblate).  Thus, the ratio  between the linear dimensions of the confined region for prolate and oblate cases is $L_{\rm pr} = 4 L_{\rm ob}$.  
  Now, we estimate the increase in the conformational free energy due to such confinement.  We consider an ideal chain confined in a region of size $L_{\rm C}$.  As the number of segments in a blob of size $L_{\rm C}$ is proportional to $L_{\rm C}^2$, a chain made of $N$ segments can be regarded as a linear chain of $N/L_{\rm C}^2$ blobs.  Therefore, the increase in the conformational free energy per chain due to the confinement is given by
\begin{equation}
  \Delta F(L_{\rm C}) = k_{\rm B} T \ln 2^{N/L_{\rm C}^2} 
                          = \frac{N k_{\rm B}T}{L_{\rm C}^2} \ln 2.
\end{equation}
Using the fact that the number of chains is inversely proportional to the chain length $N$ because of the constant volume fraction $\phi$ inside the vesicle, and the fact that there are two directions of the confinement in the prolate case, we can estimate the difference in the total conformational free energy ({\it i.e.} $\Delta F \times$ ({\rm number of chains})) between the prolate and the oblate cases as
\begin{equation}
  \frac{1}{N} \left[ 
                      2 \Delta F(L_{\rm pr}) - \Delta F(L_{\rm ob})
                    \right] 
      = - \frac{7 k_{\rm B}T \ln 2}{8 L_{\rm ob}^2} < 0.
\label{Confinement}
\end{equation}
Equation~(\ref{Confinement}) means that the conformational free energy prefers the prolate shape, which is consistent with the results in the long chain region in Fig.~\ref{FreeEnergyDifference}(a).

In Fig.~\ref{FreeEnergyDifference}(b), we show similar data as those in Fig.~\ref{FreeEnergyDifference}(a) but for the dependence on packing ratio $v$.  As $v$ becomes smaller, the polymers are more strongly confined.  Above simple consideration suggests that the prolate shape will be more and more stable than the oblate one when the constraint becomes stronger.  Actually, we can confirm this tendency in the behavior of the conformation entropy of polymers and total free enrgy shown in Fig.~\ref{FreeEnergyDifference}(b). 
\begin{figure}[t]
   \begin{center}
   \includegraphics[width=50mm]{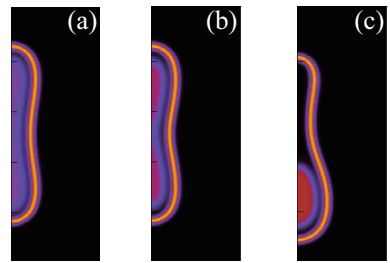}
\caption{\label{ProlateNonZeroChi} Distributions of the membrane and the polymers are shown for the case $N=100$, $v = 0.5$ and $\phi = 0.1$.  The interaction parameter $\chi$ is (a) $0.6$ (b) $0.65$ and (c) $0.7$, respectively.}
  \end{center}
\end{figure}

Finally, we consider the case where there is a repulsive interaction between the polymer segment and the solvent molecule ($\chi > 0$).
In Fig.~\ref{ProlateNonZeroChi}, we show distributions of the membrane and the polymers for the case with $N = 100$, $v = 0.5$ and $\phi = 0.1$.  The interaction parameter between the polymer segment and the solvent $\chi$ is (a) $\chi = 0.6$, (b) $\chi = 0.65$ and (c) $\chi = 0.7$, respectively.
Compared to the athermal case ($\chi = 0$) in Fig.~\ref{VesicleShapeChi0}, the polymers distribute inhomegeneously forming a depletion layer near the membrane.  While the vesicle shows a symmetric shape for a smaller value of the $\chi$-parameter (Figs.~\ref{ProlateNonZeroChi}(a) and (b)), the membrane shape becomes asymmetric for a larger value of $\chi$ ($\chi = 0.7$ in Fig.~\ref{ProlateNonZeroChi}(c)) due to an asymmetric distribution of the polymers inside the vesicle.

As a conclusion, we introduced a new field theoretic model for a vesicle that encloses polymers.  With this model, we succeeded in calculating the equilibrium shape deformation of the vesicle induced by the polymers.  This technique has a wide variety of extensions and applications such as the fusion and fission of the membrane by introducing the Gaussian curvature into the model.

\acknowledgments
The authors thank Q.Du and I.Takagi for fruitful discussions.  The present study is inspired by a collaboration of one of the authors (TK) with K.N.Yaegashi, M.Imai and N.Urakami.  The present study is supported by Grant-in-Aid for Scientific Research on Priority Area ``Soft Matter Physics'' from the Ministry of Education, Culture, Sports, Science, and Technology of Japan, and Global COE Program at Tohoku University.


\begin{thebibliography}{99}
  \bibitem{Cell}   
    \Name{B.~Alberts {\it et al.}}
    \Book{Molecular Biology of the Cell, 4th ed.}
    \Publ{Garland Science, New York}
    \Year{2002}.

  \bibitem{Banat}
    \Name{L.M.~Banat, R.S.~Makkar \and S.S.~Cameotra}
    \REVIEW{Appl. Microbiol. Biotechnol.}{53}{2000}{495}.

  \bibitem{DDS}
    \Name{T.M.~Allen, {\it et al.}}
    \REVIEW{Science}{303}{2004}{1814}.

  \bibitem{Nakaya}
    \Name{K.~Nakaya, M.~Imai, S.~Komura, T.~Kawakatsu 
           \and N.~Urakami}
    \REVIEW{Europhys. Lett.}{71}{2005}{494}.  

  \bibitem{MolecularModelMembrane}
    \Name{M.~Laradji \and P.B.S.~Kumar}
    \REVIEW{J. Chem. Phys.}{123}{2005}{224902}.

  \bibitem{Urakami} 
    \Name{T.Kurokawa, N.Urakami, K.N.Yaegashi, M.Imai \and T.Yamamoto}
    \Book{private communication}.

  \bibitem{MeshModelMembrane}
    \Name{H.~Noguchi \and G.~Gompper}
    \REVIEW{Proc. Nat. Acad. Sci.}{102}{2005}{14159}.

  \bibitem{AllenTildesley}
    \Name{M.P.~Allen \and D.J.~Tildesley}
    \Book{Computer Simulation of Liquids}
    \Publ{Oxford University Press, Oxford}
    \Year{1987}.

  \bibitem{SCF}
    \Name{T.~Kawakatsu}
    \Book{Statistical Physics of Polymers}
    \Publ{Springer-Verlag, Berlin}
    \Year{2004}.

  \bibitem{MatsenSchick}
    \Name{M.W.~Matsen \and M.~Schick}
    \REVIEW{Phys. Rev. Lett.}{72}{1994}{2660}.

  \bibitem{Du}
    \Name{Q.~Du, C.C.~ Liu \and X~ Wang}
    \REVIEW{J. Comput. Phys.}{198}{2004}{450}.

  \bibitem{Campelo}
    \Name{F.~Campelo \and A.~Hern$\acute{\rm a}$ndez-Machado}
    \REVIEW{Eur. Phys. J. E}{20}{2006}{37}.

  \bibitem{Helfrich}
    \Name{W.~Helfrich}
    \REVIEW{Z. Naturforsch.}{28 c}{1973}{693}.

  \bibitem{Seifert}
    \Name{U.~Seifert}
    \REVIEW{Adv. Phys.}{46}{1997}{13}.

  \bibitem{Du_PrivateCommunication}
    \Name{The integrand of the PF free energy eq.(\ref{PF_Free_Energy}) is a square of the variation of Ginzburg-Landau (GL) free energy. 
As the GL free energy of an interface system is proportional to the interfacial area, its variation corresponds to the mean curvature of the interface.  Thus, eq.(\ref{PF_Free_Energy}) can be identified with the Helfrich's bending energy.} 

  \bibitem{VolumeConstraint}
    \Name{In the actual PF simulation, the discontinuous condition 
$\int_{\psi({\bf r}) > 0} d{\bf r}$ is replaced by a smoothed function 
$\int d{\bf r} (1/2)\left\{ 1 + \tanh \alpha \psi({\bf r}) \right\}$ with a sufficiently large constant $\alpha$.}  
\end{thebibliography}
\end{document}